\documentclass[12pt,paper]{article} 
\usepackage{amsmath}
\usepackage{amssymb} 
\usepackage{graphics} 

 \def\C{{ \mathbb C}}  \def\Z{{
    \mathbb Z}}   \def\P{{
    \mathbb P}}
    \def\tr{\operatorname{tr}}

 \def\tr{{\rm tr\, }}

\def\id{\protect{{1 \kern-.28em {\rm l}}}}

\newcommand{\be}{\begin{equation}} \newcommand{\ee}{\end{equation}}
\newcommand{\bea}{\begin{eqnarray}} \newcommand{\eea}{\end{eqnarray}}
\newcommand{\beann}{\begin{eqnarray*}}
  \newcommand{\eeann}{\end{eqnarray*}}
\newcommand{\bfig}{\begin{figure}} \newcommand{\efig}{\end{figure}}
\newcommand{\nn}{\nonumber}
\newcommand{\ba}{\begin{array}}\newcommand{\ea}{\end{array}}

\newtheorem{Proposition}{Proposition}[section]

\newtheorem{Theorem}{Theorem}[section]
\newtheorem{Lemma}{Lemma}[section]
\newtheorem{Corrolary}{Corrolary}[section]

\newcommand{\bp}{\begin{Proposition}}
  \newcommand{\ep}{\end{Proposition}}
\newcommand{\bt}{\begin{Theorem}} \newcommand{\et}{\end{Theorem}}
\newcommand{\bl}{\begin{Lemma}} \newcommand{\el}{\end{Lemma}}
\newcommand{\bc}{\begin{Corrolary}} \newcommand{\ec}{\end{Corrolary}}

   \def\ep{\eps}

\begin{document}

\begin{titlepage}
\begin{flushright}
IFT-UAM/CSIC-03-41\\
HU-EP-03/73
\end{flushright}
\vspace{1cm}
\begin{center}{\Large\bf
On $Sp(0)$ factors and orientifolds}\\ \vspace{1.5cm}
{\large 
K. Landsteiner$^1$, 
C. I. Lazaroiu$^2$  }\\
\vspace{1cm}
$^1${\it Insitituto de F\'\i sica Te\'orica,} \\
{\it C-XVI Universidad Autonoma de Madrid, }\\
{\it 28049 Madrid,Spain}\\
{\tt Karl.Landsteiner@uam.es} \\ \vspace{1cm}
$^2${\it Humboldt Universit\"at zu Berlin}\\
{\it Newtonstrasse 15, 12489 Berlin-Adlershof, Germany}\\
{\tt calin@physik.hu-berlin.de}\\
\vspace{2cm}
\bf Abstract
\end{center}
\noindent We discuss the geometric engineering of $SO/Sp$ gauge theories
with symmetric or antisymmetric tensor matter and show that the
`mysterious' rank zero gauge group factors observed by a few authors 
can be traced back to the effects of an orientifold which survives the
geometric transition. By mapping the Konishi constraints of such
models to those of the $U(N)$ theory with adjoint matter, we show that
the required shifts in the ranks of the unbroken gauge group
components is due to the flux contribution of the orientifold after
the transition.

\end{titlepage}

\noindent{\bf Introduction} \\

It was recently pointed out \cite{Cachazo} that the Konishi constraints
\cite{cdsw} 
of the
$Sp(N)$ theory with antisymmetric matter and a tree-level superpotential
of degree $d+1$ can be mapped to those of the
$U(N+2d)$ theory with adjoint matter. 
By noticing that this relation involves a shift in the rank of the
components of the unbroken gauge groups, it was shown 
\cite{Cachazo, Matone} that an apparent discrepancy found in \cite{KS}
and further explored in \cite{vafa_sp} can be removed by relaxing an
unwarranted assumption. Moreover, it was speculated 
that the somewhat mysterious $Sp(0)$ factors involved in this relation 
originate in the IIB realization of such field theories.

In the present note, we show that the observations of
\cite{Cachazo} have a simple interpretation in the geometric 
engineering of such models, and they admit an obvious
generalization. By considering the four $SO/Sp$
theories with symmetric or antisymmetric matter, we show that their 
IIB realization involves a $\Z_2$ orientifold of an $A_1$ fibration. 
As in \cite{llt1, llt2}, we find that the orientifold 5-plane involved in
this construction survives the geometric transition of 
\cite{vafa, civ, Cachazo_Vafa, Cachazo_Vafa_more}. This allows us to 
show that the phenomena observed in \cite{Cachazo} are due to the
flux contribution of this orientifold after the transition. Moreover,
we show that the Konishi constraints of all four models can be mapped 
to those of a theory with unitary gauge group and adjoint matter, and that this
map amounts to replacing the orientifold by its flux contribution. This gives an elementary
explanation of the relation found in \cite{Cachazo}. 

We shall be interested in ${\cal N}=1$ gauge theories with gauge group
$G=SO(N)$ or $Sp(N)$\footnote{We use conventions in which $N$ is even
  for $Sp(N)$.} and a single chiral
superfield $X$ with $X^T=\epsilon X$ and $\epsilon =\pm 1$ for 
the symmetric or antisymmetric representation. The gauge
transformation is:
\be
\label{field_gauge}
X\rightarrow UXU^T~~
\ee
with $U$ valued in $G$. Consider the tree-level superpotential: 
\begin{equation}
\label{W_tree}
W_{tree}=\tr \left[ W(\Phi)\right] 
\end{equation} 
where $\Phi=X$ for $G=SO(N)$ and $\Phi=XJ$ for $G=Sp(N)$, where 
$J=\left[\ba{cc} 0 & 1_{N/2}\\-1_{N/2} & 0\ea \right]$. 
Here: 
\begin{equation}
\label{W}
W(z)=\sum_{j=1}^{d+1}{\frac{t_j}{j}z^j} 
\end{equation} 
is a complex polynomial of degree $d+1$. Throughout this paper, we
assume that $W'(z)$ has simple zeroes. Since $U^{-T}JU^{-1}=J$ for
$U\in Sp(N)$ and $U^T=U^{-1}$ for $U\in SO(N)$, the
field $\Phi$ always transforms as $\Phi\rightarrow U\Phi U^{-1}$.
In particular, $\Phi$ is in the adjoint representation for the
antisymmetric representation of $SO(N)$ and the symmetric
representation of $Sp(N)$. In these cases, we can assume that $W$ 
is an even polynomial since only even powers of $\Phi$ contribute to
(\ref{W_tree}).

\vspace{1cm}

\noindent{\bf Geometric engineering}\\

\noindent To find the IIB realization of our models, we distinguish
the cases: 
\begin{itemize}
\item[(A)] $SO(N)$ with symmetric matter or $Sp(N)$ with antisymmetric matter
\item[(B)] $SO(N)$ with antisymmetric matter or $Sp(N)$ with symmetric matter
\end{itemize}
The engineering  of $(A)$ was given in \cite{llt2}
and that of $(B)$ was discussed in \cite{llt1} \footnote{Case $(B)$ had already been 
engineered in \cite{eot,fo}, but in a framework different from the one
we shall find useful here. In the approach of \cite{eot,fo}, the 
IIA T-dual involves an orientifold 4-plane. In this paper, we 
use the construction of \cite{llt1}, whose IIA dual involves an orientifold 6-plane.
The relation between the two realizations is discussed in \cite{llt1}.}
. 
In both cases, we start with the singular $A_1$ fibration given by:
\begin{equation}
\label{X_10}
X_{0}:~~xy=(u-W'(z))(u+W'(z))~~,  \end{equation} which admits the
two-section: \begin{equation}
\label{Sigma_10}
\Sigma_{0}:~~x=y=0,~~(u-W'(z))(u+W'(z))=0~~. 
\end{equation}
This is a union of two rational curves which intersect at the critical
points $z_j$ of $W$. Since $W$ is even in case $(B)$, we let $d=2n+1$ 
and take $j=-n\dots n$. For case $(A)$ we take
$j=1\dots d$. 

The resolution ${\hat X}$ can be described as the complete
intersection: 
\begin{eqnarray}
\label{cicy}
  \beta ( u - W'(z)) &=& \alpha x \nn\\
  \alpha (u +W'(z)) &=& \beta y \\
  (u-W'(z))(u+W'(z)) &=& xy~~\nn \end{eqnarray} in the ambient space
$\P^1[\alpha,\beta]\times \C^4[z, u, x, y]$.  The exceptional $\P^1$'s
are denoted by $D_j$ and sit above the singular points of $X_{0}$, which are determined by
$x=y=u=0$ and $z=z_j$.  The resolved space
admits the $U(1)$ action: \begin{equation}
\label{U1red}
([\alpha, \beta], z, u, x, y) \longrightarrow ([e^{-i\theta} \alpha,
\beta], z, u,e^{i\theta} x, e^{-i\theta}y)~~. \end{equation}

For the two cases, consider the holomorphic $\Z_2$
actions\footnote{These square to the identity since
  $[-\alpha,-\beta]=[\alpha,\beta]$ in $\P^1$.}:
\bea
\label{red_or}
{\hat k}_A:~~([\alpha,\beta],z,u,x,y) &\longrightarrow& ([-\beta,
\alpha], z, -u, y, x)~~\nn\\
{\hat k}_B:~~([\alpha,\beta],z,u,x,y) &\longrightarrow& ([-\beta,
\alpha], -z, u, -y, -x)~~,
\eea 
which obviously preserve (\ref{cicy}) (remember that $W'(-z)=-W'(z)$ in case $(B)$).
The first symmetry preserves each exceptional curve, while the second
preserves $D_0$ while exchanging $D_j$ with $D_{-j}$. 

These symmetries project to the following
involutions of $X_{0}$: 
\bea
\label{red_or_proj}
\kappa_{A0}:~~(z,u,x,y) &\longrightarrow& (z, -u, y,x)~~\\
\kappa_{B0}:~~(z,u,x,y) &\longrightarrow& (-z, u, -y,-x)~~\nn
\eea
whose fixed point sets are given by: 
\bea
\label{Olocus}
& & O_{A0}:~x=y,~~u=0,~~x^2+W'(z)^2=0~~\nn\\
& & O_{B0}:~x=-y,~~z=0,~~x^2+u^2=0 .\nn
\eea           
The fixed point loci of (\ref{red_or}) are:
\bea
& & {\hat O}_A:~~x-y=u=x^2+W'(z)^2=0~~,~~\frac{\alpha}{\beta}=\pm i~~\\
& & {\hat O}_B:~~x+y=z=x^2+u^2=0~~~~~~~,~~\frac{\alpha}{\beta}=\pm i
\eea
We shall use the geometric symmetries (\ref{red_or}) to define orientifolds
of our IIB background upon combing them with worldsheet parity
reversal. More precisely, we choose the orientifold projections such
that ${\hat O}_A$ corresponds to an $O_5^{-\epsilon}$ plane and ${\hat
  O}_B$ corresponds to an $O_5^{+\epsilon}$ plane.

It is not hard to check that this construction engineers our
theories. 
The matter content can be recovered geometrically or by a fractional
brane construction. More directly, one can follow the approach of
\cite{OT, llt1, llt2} by using T-duality to map our
background to the Hanany-Witten realizations of these models.  
\vspace{1cm}

\noindent{\bf Dual configurations}\\

To extract the T-dual Hanany-Witten systems, we use a local
description valid on a subset $\tilde X \subset \hat X$.  
This is given by two copies $U_0$ and $U_1$ of $\C^3$ with
coordinates $(x_i, u_i, z_i)$ ($i=0,1$) which are glued 
according to: \begin{equation} (x_1, u_1, z_1) = (\frac{1}{u_0}, x_0
  u_0^2 -2 W'(z_0) u_0, z_0)~~. \end{equation} 
The resolution map is given by:
\begin{eqnarray}
(z,u,x,y) =& ( z_0, x_0 u_0 - W'(z_0), x_0 , u_0(x_0 u_0 - 2 W'(z_0)) ) \,,\\
          =& ( z_1, x_1 u_1 + W'(z_1), x_1(x_1 u_1 + 2 W'(z_1)), u_1
          )\nn ~~,
\end{eqnarray}
while the $U(1)$ action (\ref{U1red}) takes the form: 
\begin{equation}
\label{U1red_local}
(z_i, u_i, x_i ) \longrightarrow (z_i, e^{-i\theta}u_i,e^{i\theta}x_i)~~. 
\end{equation} 
Its fixed point set is the union of
rational curves $x_0=u_0=0$ and $x_1=u_1=0$. This action stabilizes the
exceptional curves $D_j:~x_0=u_1=z-z_j=0$.

The Hanany-Witten construction results by T-duality with respect
to the circle orbits of this action. Following \cite{llt1}, we 
use the following ansatz for the T-dual coordinates:
\bea
w &:=& x^4 + i x^5 = x_0 u_0 - W'(z_0)=x_1 u_1 + W'(z_1)~~,\nn \\
x^6 &=&\frac{1}{2}(|x_1|^2-|u_0|^2)~~,~~\nn \\
z&=&x^8+ix^9~~~~~~~~~~~~~~~\nn
\label{base_coords_a1}
\eea
together with the periodic coordinate $x^7$ along the orbits of
(\ref{U1red_local}). 

Expressing the fixed point set of (\ref{U1red_local}) in these
coordinates, we find that the dual background
contains two NS5-branes ${\cal N}_0$ and ${\cal N}_1$ sitting at:
\bea
& & {\cal N}_0:~~w=-W'(z)~~,~~x^6=+\infty\nn\\
& & {\cal
    N}_1:~~w=+W'(z)~~,~~x^6=-\infty~~. \nn
\eea 
We also have D4-branes ${\cal D}_j$ stretching between the NS5-branes at 
$z=z_j$.

\

\begin{figure}[hbtp]
\begin{center}
  \scalebox{0.6}{\input{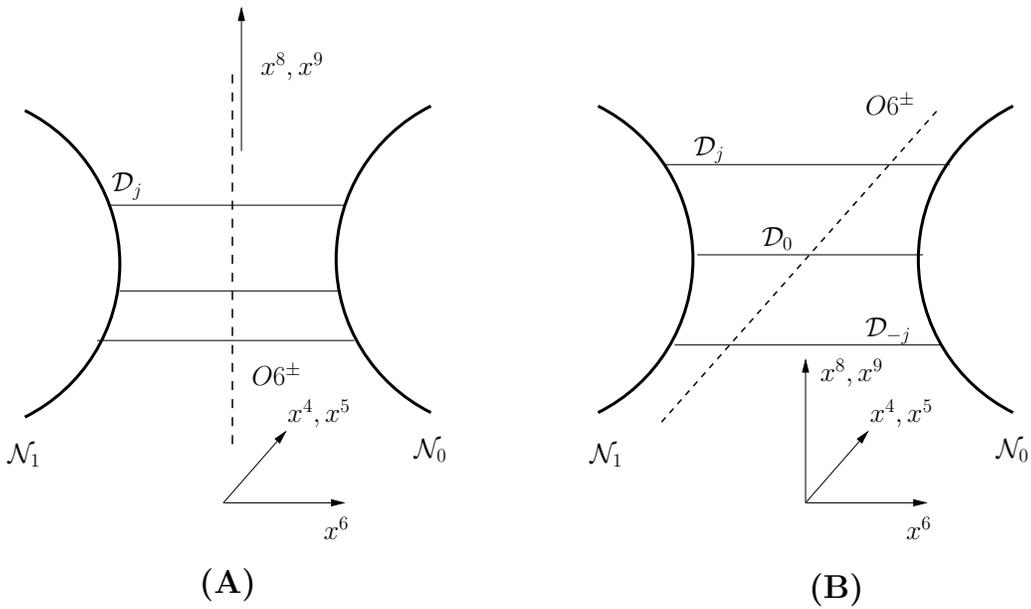}}
\end{center}
\caption{\footnotesize Brane configuration for the $SO(N)/Sp(N)$ 
theories with symmetric or antisymmetric matter. The outer NS5-branes are
  bent in the directions $x^4$ and $x^5$, which cannot be shown
  properly in this two-dimensional figure. The orientifold plane has charge $-4\epsilon$
in case $(A)$ and $+4\epsilon$ in case $(B)$.} 
\label{O6plane}
\end{figure}

\

The orientifolds (\ref{red_or}) act as:
\bea
\label{red_local}
(z_0, x_0 , u_0) & \stackrel{{\hat \kappa}_A}{\longleftrightarrow}&
(z_1, u_1, -x_1)~~\nn\\  
(z_0, x_0 , u_0) & \stackrel{{\hat \kappa}_B}{\longleftrightarrow}&
(-z_1, -u_1, -x_1)~~ 
\eea 
In the first case, the  fixed point set is ${\hat
  O}_A:~u_0^2+1=x_0+W'(z)u_0=0$.  In the second case, it is 
${\hat O}_B:~u_0^2+1=z=0$. Both of these are unions of two disjoint rational curves. 
The IIA orientifold action is:
\bea
& &(A): x^6 \rightarrow -x^6~~,~~z\rightarrow z~~,~~ w\rightarrow -w \nn \\
& &(B): x^6 \rightarrow -x^6~~,~~z\rightarrow -z~~,~~ w\rightarrow w~~.
\eea
Using (\ref{base_coords_a1}) we find that under T-duality these loci map to
O6-planes sitting at $x^4=x^5=x^6=0$ and $x^6=x^8=x^9=0$
respectively (figure \ref{O6plane}). This recovers the Hanany-Witten
realization of our models\footnote{For a detailed discussion of these constructions and further references see \cite{gk}}. 
\vspace{1cm}

\noindent{\bf Description after the geometric transition}\\

After the geometric transition of \cite{vafa,civ,Cachazo_Vafa,Cachazo_Vafa_more}, the Calabi-Yau 
space (\ref{X_10}) is deformed
to: \begin{equation}
\label{X1}
X:~~xy=u^2-W'(z)^2-f(z)~~, 
\end{equation} where $f(z)$ is a polynomial
of degree at most $d-1$.  This fibration admits the two-section: \begin{equation}
\label{Sigma_1}
\Sigma:~~x=y=0,~~u^2-W'(z)^2-f(z)=0~~. 
\end{equation}  
The D5-branes wrapping the exceptional divisors are replaced by
fluxes. Writing $W'(z)^2+f(z)=\prod_{j}{(z-a_j)(z-b_j)}$, we can
choose the cuts $I_j$ of (\ref{Sigma_1}) to connect $a_j$ and
$b_j$. We also choose a symplectic basis of cycles $A_j, B_j$ with
$A_j$ associated with the cut $I_j$. In case $(B)$ we can choose these such that $I_{-j}=-I_j$. 
In particular, we have the cut $I_0$ which passes
through the origin. 

In case $(A)$, the deformed space
(\ref{X1}) is still invariant under the $\Z_2$ action
(\ref{red_or_proj}) so the orientifold 5-plane survives the
transition. Its internal part is deformed to the irreducible curve:
\begin{equation}
\label{ora} 
O_A:~~x=y~~,~~u=0~~,~~x^2+W'(z)^2+f(z)=0~~.  
\end{equation}

In case $(B)$, the polynomial $f(z)$ must be even in order to preserve
the orientifold symmetry. Again the orientifold survives the 
transition, after which its internal part becomes:
\begin{equation}
\label{orb} 
O_B:~~x=-y~~,~~z=0~~,~~x^2+u^2-f(0)=0~~.  
\end{equation}

The Riemann surface (\ref{Sigma_1}) arises naturally in the confining
phase of the $SO(N)/Sp(N)$ theories with (anti)symmetric matter 
\cite{Alday,KRS}.
This curve can be extracted by analyzing
the generalized Konishi anomalies of such theories. 
\vspace{1cm}

\noindent{\bf Relation to generalized Konishi constraints}\\

Consider the field theory quantities $T(z)=\langle \tr
\frac{1}{z-\Phi}\rangle$ and $R(z)=\langle \tr \frac{{\cal
    W}^2}{z-\Phi}\rangle$, where ${\cal W}_\alpha$ is the superfield strength.  

\vspace{0.5cm}

\noindent{\em Case (A)}\\

Using the method of generalized Konishi anomalies, it was shown in
\cite{Alday,KRS} that $R(z)$ and $T(z)$ satisfy: 
\bea
\label{konishia}
W'R&=&\frac{1}{2}R^2-\frac{f}{2}\\
W'T&=&TR-2\epsilon R'+c~~,\nn
\eea
where $f$ and $c$ are polynomials of degree at most $d-1$.
The solution is: 
\bea
R&=&W'-u\nn\\
T&=&\frac{c}{u}-2\epsilon \frac{W''-u'}{u}={\tilde T}-\Psi
\eea
where ${\tilde T}=\frac{{\tilde c}}{u}$ with ${\tilde c}=c-2\epsilon
W''$ a polynomial of degree at most $d-1$, $\Psi=-2\epsilon \frac{u'}{u}$ 
and $u=\sqrt{(W')^2+f}$ is the appropriate branch of the spectral curve (\ref{Sigma_1}). 
The pair $(R,{\tilde T})$ satisfies the relations:
\bea
\label{konishiu}
W'R&=&\frac{1}{2}R^2-\frac{f}{2}\\
W'{\tilde T}&=&{\tilde T}R+{\tilde c}~~,\nn
\eea
which are also obeyed by the quantities $r=\langle\tr\frac{{\cal W}^2}{z-\phi} \rangle$ and
$t=\langle \tr \frac{1}{z-\phi}\rangle$ of a  theory with unitary gauge group 
and an adjoint chiral multiplet $\phi$. It is clear that $Rdz$ and ${\tilde
  T}dz$ have no poles at finite $z$ on the spectral curve
(\ref{Sigma_1}), while $\Psi dz$ has simple poles at the
branching points of $\Sigma$. 

At the branching points, $\Psi$ behaves like
$-\frac{\epsilon}{z-a_j}$ or $-\frac{\epsilon}{z-b_j}$.
 The quantity ${\cal A}=\Psi dz$ satisfies:
\footnote{Remember that ${\bar \partial}_z \frac{1}{z-a}=\pi\delta(z-a)$.}: 
\be
\label{curvature}
{\bar \partial}{\cal A}=-\epsilon
\pi \left[ 
\sum_{j=1}^d{\delta(z-a_j)}+\sum_{j=1}^d{\delta(z-b_j)}\right]d{\bar
z} d z~~.
\ee
Thus ${\cal A}$ can be viewed as the potential produced by  charges equal to $-\epsilon$
  placed at branching points of $\Sigma$. 
The `vacuum' term ${\tilde T} dz$ in ${\cal B}:=Tdz={\tilde T}dz -{\cal
  A}$ contributes fluxes through the A-cycles of $\Sigma$:
\be
\label{Nshifta}
N_j:=\oint_{A_j}{\frac{dz}{2\pi i}T}={\tilde N}_j+2\epsilon~~, 
\ee
where $\tilde N_j=\oint_{A_j}{\frac{dz}{2\pi i}{\tilde T}}$ with $+2\epsilon$ the
  contribution from $-{\cal A}$. In view of the above, 
relation (\ref{Nshifta}) maps a vacuum
of our theory  with unbroken gauge group $\prod_{j=1}^{d}{SO(N_j)}$
($\epsilon=+1$) or
  $\prod_{j=1}^{d}{Sp(N_j)}$ ($\epsilon=-1$) to a $\prod_{j=1}^d{U(N_j-2\epsilon)}$ vacuum
  of the $U(N-2\epsilon d)$ theory with adjoint matter.

It is easy to find the IIB interpretation of this map. Recall that the
orientifold survives the geometric transition,
giving an $O_5^{-\epsilon}$ plane whose internal directions wrap the curve
(\ref{ora}). This curve intersects the Riemann surface (\ref{Sigma_1})
precisely at its branching points $(z,y)=(a_j,0)$ or $(b_j,0)$, and
contributes to the flux through  the 3-cycles $S_j$ associated with
the cuts $I_j$\footnote{As in \cite{klmvw,civ}, the 
  3-cycles of $X$ can be constructed by fibering two-spheres over the
  cuts.}. This accounts for the shift by $2\epsilon$ in relation
(\ref{Nshifta}). More precisely, $N_j$ is the number of D-branes
wrapping the exceptional curves $D_j$ before the transition, while
${\tilde N}_j=N_j-2\epsilon$ is the total RR flux through the
associated 3-cycle produced after the transition. The flux contribution
$N_j$ is due to the D-brane wrapping $D_j$, which is replaced by a RR
flux during the transition, while $-2\epsilon$ is 
the the flux contribution of the $O_5^{-\epsilon}$
plane (\ref{ora})\footnote{
In our case, the orientifold 5-plane intersects the 3-cycle $S_j$ 
along a circle. The RR 3-form $H$ is not closed due to the presence of
the orientifold ($H$ has a source supported along the curve
(\ref{ora})). One can construct an $S^2$ fibration ${\cal S}$ of $X$ over the
$z$-plane whose $S^2$ fibers are themselves obtained 
by fibering circles over the intervals $I_z=[u_-(z), u_+(z)]$ 
in the $u$-plane, where $u_{\pm}(z):=\pm \sqrt{W'(z)^2+f(z)}$. The
fibers of ${\cal S}$ collapse to points for $z=a_j$ or $z=b_j$. Then the
integral of $H$ over $S_j$ equals the integral of
${\tilde {\cal A}}$ over $A_j$, where ${\tilde {\cal A}}$ is a
(non-meromorphic) one-form on $\Sigma$
obtained from $\frac{1}{2}H$ by `push-forward' along the $S^2$ fibration ${\cal
  S}$. As in \cite{csw2} ${\tilde {\cal A}}$ has integral periods but differs
from ${\cal A}$ by a one-form whose periods
vanish on-shell. }.

Thus the shift observed in
\cite{Cachazo} is explained by the presence of an O5 plane after the
geometric transition. Moreover, it is clear that the map
$(R,T)\rightarrow  (R, {\tilde T})$ to the $U(N-2\epsilon d)$ theory amounts to
replacing the orientifold by its flux contribution, i.e. 
considering the IIB theory with the same total RR fluxes and on the same
geometry (\ref{X1}), but without the orientifold plane (\ref{ora}). 
The latter IIB background is well-known to engineer the $U(N-2\epsilon
d)$  theory with adjoint matter. 
Hence the map of \cite{Cachazo} has an
elementary interpretation in geometric
engineering\footnote{Other relations of this type were considered in 
\cite{llt1}, where they
  were shown to have similarly straightforward interpretations.}. 

Of course this map only
  refers to matching of the associated Konishi constraints, and should
  not be taken at face value regarding other quantities of
  physical interest. For case of an $SO(N)$ group with symmetric matter (i.e. $\epsilon = +1$)
we can have  $\tilde N_j < 0$ for some $j$. This simply means that the total 
flux through the associated 3-cycle is allowed to become
negative. 
This is of course purely formal in the context of the $U(N-2\epsilon d)$
theory, and only receives its proper physical
interpretation once one considers the orientifold, thereby recovering
the $SO/Sp$ model.

For $N_j=2$ one finds that an $SO(2)$ factor group is mapped to a
$U(0)$ factor. 
In the engineering of the $U(N-2\epsilon d)$ model, 
this means that there are no branes wrapping the corresponding $\P^1$ 
before the transition, and no RR flux through the associated 3-cycle
after the transition. In particular, one can keep this cycle
collapsed, in which case the associated cut of
the spectral curve (\ref{Sigma_1}) is reduced to a double point.
Nevertheless, it is clear that the period of $T$ does not vanish in
this limit because of the flux contribution of the orientifold, which
passes through this double point. 
This behavior of the $SO$ theory with symmetric matter was conjectured in \cite{Cachazo}.
We note that similar effects were already 
found in \cite{us} for the more complicated case of
$U(N)$ theories with adjoint and symmetric or antisymmetric matter, and
explained in \cite{llt1} in terms of an orientifold which survives the
geometric transition. 

\vspace{0.5cm}
%\subsection{Case B}

\noindent {\em Case (B)}\\

It was shown in \cite{Alday,KRS} that $R(z)$ and $T(z)$ satisfy: 
\bea
\label{konishib}
W'R&=&\frac{1}{2}R^2-\frac{f}{2}\\
W'T&=&TR+\frac{2\epsilon}{z} R+c~~,\nn
\eea
where $f$ and $c$ are polynomials of degree at most $d-1$. The solution is: 
\bea
R&=&W'-u\nn\\
T&=&\frac{c}{u}+\frac{2\epsilon}{z}\left[\frac{W'}{u}-1\right]={\tilde T}-\Psi
\eea
where ${\tilde T}=\frac{{\tilde c}}{u}$ with ${\tilde c}=c+2\epsilon
\frac{W'}{u}$ a polynomial of degree at most $d-1=2n$ (remember that $W'$ is
odd !) and $\Psi=+\frac{2 \epsilon}{z}$. 
The pair $(R,{\tilde T})$ satisfies the relations (\ref{konishiu}) of a theory with
unitary gauge group and an adjoint chiral multiplet. We have: 
\be
{\bar \partial}\Psi=2\pi \epsilon \delta(z)d{\bar z}~~
\ee
and:
\be
\label{Nshiftb}
N_j:=\oint_{A_j}{\frac{dz}{2\pi i}T}={\tilde N}_j~~({\rm for}~j\neq 0)~~,~~ 
N_0:=\oint_{A_0}{\frac{dz}{2\pi i}T}={\tilde N}_0-2\epsilon~~
\ee
with ${\tilde N}_j$ the contributions from ${\tilde T}$. We have
$N_{-j}=N_j$ for all $j$. 

The IIB interpretation is as before. After the geometric transition, 
the $O_5^{+\epsilon}$ plane (\ref{orb}) pierces the spectral curve
(\ref{Sigma_1}) in the two points $u=\pm
\sqrt{f(0)}$ sitting above $z=0$.
It contributes $+2\epsilon$ to the RR flux ${\tilde N}_0$ through the associated
$S^3$ cycle in $X$, leading to the relation ${\tilde N}_0=N_0 +2\epsilon$.
This allows us to identify a vacuum
of our theory  with unbroken gauge group $SO(N_0)\times
\prod_{j=1}^{n}{SU(N_j)}$ ($\epsilon=-1$) or
  $Sp(N_0)\times \prod_{j=1}^{n}{SU(N_j)}$ ($\epsilon=+1$) with an 
$SU(N_0+2\epsilon)\times \prod_{j=1}^n{(U(N_j)\times U(N_j))}$ vacuum of the 
$U(N+2\epsilon)$ theory with adjoint matter.  
Again this identification is only formal in the case $\epsilon=-1$
(i.e. $SO(N)$ with antisymmetric matter) and
$N_0=0$. 

\vspace{0.5cm}

\noindent{\bf Conclusions}\\

We considered the geometric engineering and T-dual Hanany-Witten
realizations of four field theories, namely $SO(N)$ with symmetric or
antisymmetric matter and $Sp(N)$ with symmetric or antisymmetric
matter. As in \cite{llt1, llt2}, we found that the IIB realization of such
models involves a $\Z_2$  orientifold which survives the geometric 
transition of \cite{vafa, civ, Cachazo_Vafa, Cachazo_Vafa_more} and
therefore contributes to the effective superpotential and fluxes.
Following \cite{Cachazo}, we extracted a relation between the Konishi 
constraints of such theories and those of the $U(\tilde N)$ field theory
with adjoint matter, where $\tilde N = N-2\epsilon d$ for SO/Sp with
symmetric ($\epsilon=1$)/antisymmetric ($\epsilon=-1$)matter and 
$\tilde N = N+2\epsilon$ for SO/Sp with antisymmetric/symmetric matter.
Its interpretation in geometric
engineering amounts to the trivial operation of replacing the
orientifold 5-plane by its flux contribution.

The fact that the orientifold  
contributes to the flux through various 3-cycles after the transition
is responsible for the phenomena discussed in
\cite{Cachazo} and formalized in \cite{Matone}. In particular, it
gives an elementary explanation of the rank shifts required by  the relation
with the $U(\tilde N)$ theory. It also recovers and generalizes the role of 
$Sp(0)$ factors in the
$Sp(N)$ theory with antisymmetric matter. 
For the particular case of the $SO(N)$ theory with symmetric matter, 
we confirmed the conjecture of \cite{Cachazo} that $T(z)$ can have non-vanishing period
even if the associated branch cut on the Riemann surface is collapsed
to a double point. As in \cite{llt1}, we find
that simple operations in geometric engineering account for
non-obvious relations between strongly coupled field theories.

\vspace{0.5cm}

\noindent{\bf Acknowledgments:} This work was supported by DFG grant KL1070/2-1.

\end{document}